\documentstyle[12pt]{article}
\newcommand{\nc}{\newcommand}
\nc{\postscript}[2]
{\setlength{\epsfxsize}{#2\hsize}\centerline{\epsfbox{#1}}}
\nc{\bg}{B. Grzadkowski}
\nc{\non}{\nonumber}
\def\dps{\displaystyle}
\def\mib#1{\mbox{\boldmath $#1$}}

\def\bra#1{\langle #1 |} \def\ket#1{|#1\rangle}
\def\vev#1{\langle #1\rangle}

\nc{\barx}{\bar{x}}\nc{\pbarn}{\;\hbox {pb}}\nc{\fbarn}{\;\hbox {fb}}
\nc{\hc}{\hbox {h.c.}} \nc{\re}{\hbox {Re}} 
\nc{\mev}{\hbox {MeV}} \nc{\gev}{\;\hbox {GeV}}
\def\gesim{\lower0.5ex\hbox{$\:\buildrel >\over\sim\:$}}
\def\lesim{\lower0.5ex\hbox{$\:\buildrel <\over\sim\:$}}
\nc{\prd}[3]{{\it Phys.\ Rev.}\ {{\bf D{#1}} (#2), #3}}
\nc{\prl}[3]{{\it Phys.\ Rev.\ Lett.}\ {{\bf {#1}} (#2), #3}}
\nc{\plb}[3]{{\it Phys.\ Lett.}\ {{\bf B{#1}} (#2), #3}}
\nc{\npb}[3]{{\it Nucl.\ Phys.}\ {{\bf B{#1}} (#2), #3}}
\nc{\ptp}[3]{{\it Prog.\ Theor.\ Phys.}\ {{\bf {#1}} (#2), #3}}
\nc{\zfp}[3]{{\it Z.\ Phys.}\ {{\bf C{#1}} (#2), #3}}
\nc{\epj}[3]{{\it Eur.\ Phys.\ J.}\ {{\bf C{#1}} (#2), #3}}
\nc{\mpla}[3]{{\it Mod.\ Phys.\ Lett.}\ {{\bf A{#1}} (#2), #3}}
\nc{\rmp}[3]{{\it Rev.\ Mod.\ Phys.}\ {{\bf {#1}} (#2), #3}}
\nc{\ijmpa}[3]{{\it Int.\ J.\ of\ Mod.\ Phys.}\
               {{\bf A{#1}} (#2), #3}}
\nc{\ttbar}{t\bar{t}}         \nc{\bbbar}{b\bar{b}}
\nc{\tanb}{\tan \beta}        \nc{\twbdec}{t\to W^+ b}
\nc{\tbwbdec}{\bar{t}\to W^- \bar{b}}
\nc{\epem}{e^+e^-}            \nc{\eett}{\epem \to \ttbar}
\nc{\sigeett}{\sigma_{e\bar{e}\to\ttbar}}
\nc{\wpwm}{W^+W^-}            \nc{\tbar}{\bar{t}}
\nc{\bbar}{\bar{b}}           \nc{\wpp}{W^+}
\nc{\mt}{m_t}    \nc{\mts}{m_t^2}   \nc{\mw}{m_W}    \nc{\mws}{m_W^2}
\nc{\mz}{m_Z}    \nc{\mzs}{m_Z^2}
\nc{\ttbardec}{\ttbar \to W^+W^-\bbbar}
\nc{\wwbb}{W^+W^-\bbbar}      \nc{\sm}{SM}
\nc{\cw}{\cos\theta_W}        \nc{\sw}{\sin\theta_W}
\nc{\sws}{\sin^2\theta_W}     \nc{\sig}{\sigma_{tot}}
\nc{\lp}{{\ell}^+}              \nc{\lm}{{\ell}^-}
\nc{\epsl}{\epsilon_L}        \nc{\cp}{C\!P}
\nc{\splus}{s_+}       \nc{\smin}{s_-}        \nc{\eps}{\epsilon}
\nc{\psp}{Ps_+}        \nc{\psm}{Ps_-}        \nc{\lsp}{ls_+}
\nc{\lsm}{ls_-}        \nc{\sss}{s_+s_-}      \nc{\m}{m_t}
\nc{\mq}{m_t^2}        \nc{\mr}{\frac{1}{\m}} \nc{\av}{A_{\gamma}}
\nc{\bv}{B_{\gamma}}   \nc{\az}{A_Z}          \nc{\bz}{B_Z}
\nc{\avs}{A_{\gamma}^2}\nc{\azs}{A_Z^2}       \nc{\bzs}{B_Z^2}
\nc{\dav}{\delta \! A_{\gamma}}   \nc{\dbv}{\delta \! B_{\gamma}}
\nc{\dcv}{\delta C_{\gamma}}      \nc{\ddv}{\delta \! D_{\gamma}}
\nc{\daz}{\delta \! A_Z}          \nc{\dbz}{\delta \! B_Z}
\nc{\dcz}{\delta C_Z}             \nc{\ddz}{\delta \! D_Z}
\nc{\dev}{\delta \! E_{\gamma}}   \nc{\dez}{\delta \! E_Z}
\nc{\dfv}{\delta \! F_{\gamma}}   \nc{\dfz}{\delta \! F_Z}
\nc{\rdav}{{\rm Re}(\delta \! A_{\gamma}) \:}
\nc{\rdbv}{{\rm Re}(\delta \! B_{\gamma}) \:}
\nc{\rdcv}{{\rm Re}(\delta C_{\gamma}) \:}
\nc{\rddv}{{\rm Re}(\delta \! D_{\gamma}) \:}
\nc{\rdaz}{{\rm Re}(\delta \! A_Z) \:}
\nc{\rdbz}{{\rm Re}(\delta \! B_Z) \:}
\nc{\rdcz}{{\rm Re}(\delta C_Z) \:}
\nc{\rddz}{{\rm Re}(\delta \! D_Z) \:}
\nc{\idav}{{\rm Im}(\delta \! A_{\gamma}) \:}
\nc{\idbv}{{\rm Im}(\delta \! B_{\gamma}) \:}
\nc{\idcv}{{\rm Im}(\delta C_{\gamma}) \:}
\nc{\iddv}{{\rm Im}(\delta \! D_{\gamma}) \:}
\nc{\idaz}{{\rm Im}(\delta \! A_Z) \:}
\nc{\idbz}{{\rm Im}(\delta \! B_Z) \:}
\nc{\idcz}{{\rm Im}(\delta C_Z) \:}
\nc{\iddz}{{\rm Im}(\delta \! D_Z) \:}
\nc{\cz}{(1+v_e^2)d\:\!'^2}         \nc{\ci}{v_ed\:\!'}
\nc{\ccz}{v_ed\:\!'^2}              \nc{\cci}{d\:\!'}
\nc{\lspace}{\;\;\;\;\;\;\;\;\;\;}  \nc{\llspace}{\lspace \lspace}
\nc{\beq}{\begin{equation}}   \nc{\eeq}{\end{equation}}
\nc{\bea}{\begin{eqnarray}}   \nc{\eea}{\end{eqnarray}}
\nc{\baa}{\begin{array}}      \nc{\eaa}{\end{array}}
\nc{\bit}{\begin{itemize}}    \nc{\eit}{\end{itemize}}
\nc{\ben}{\begin{enumerate}}  \nc{\een}{\end{enumerate}}
\nc{\ocal}{{\cal O}}
\begin{document}
\pagestyle{empty} \setlength{\footskip}{2.0cm}
\setlength{\oddsidemargin}{0.5cm} \setlength{\evensidemargin}{0.5cm}
\renewcommand{\thepage}{-- \arabic{page} --}
\def\mib#1{\mbox{\boldmath $#1$}}
\def\bra#1{\langle #1 |}      \def\ket#1{|#1\rangle}
\def\vev#1{\langle #1\rangle} \def\dps{\displaystyle}
\nc{\tb}{\stackrel{{\scriptscriptstyle (-)}}{t}}
\nc{\bb}{\stackrel{{\scriptscriptstyle (-)}}{b}}
\nc{\fb}{\stackrel{{\scriptscriptstyle (-)}}{f}}
\nc{\pp}{\gamma \gamma}
\nc{\pptt}{\pp \to \ttbar}
   \def\thebibliography#1{\centerline{REFERENCES}
     \list{[\arabic{enumi}]}{\settowidth\labelwidth{[#1]}\leftmargin
     \labelwidth\advance\leftmargin\labelsep\usecounter{enumi}}
     \def\newblock{\hskip .11em plus .33em minus -.07em}\sloppy
     \clubpenalty4000\widowpenalty4000\sfcode`\.=1000\relax}\let
     \endthebibliography=\endlist
   \def\sec#1{\addtocounter{section}{1}\section*{\hspace*{-0.72cm}
     \normalsize\bf\arabic{section}.$\;$#1}\vspace*{-0.3cm}}
\vspace*{-0.7cm}
\begin{flushright}
$\vcenter{
\hbox{TOKUSHIMA Report}
\hbox{(hep-ph/0407319)}
}$
\end{flushright}

\vskip 1.7cm
\renewcommand{\thefootnote}{*}

\begin{center}
{\large\bf What Will LC Tell Us on Top/QCD ?}
\footnote{Summary Talk of Top/QCD session at
{\it International Workshop on Physics and Experiments with
Future Electron-Positron Linear Colliders (LCWS2004)},
April 19-23, 2004, Le Carre des Sciences, Paris, France.}
\end{center}

\vspace*{1cm}
\renewcommand{\thefootnote}{*)}
\begin{center}
{\sc Zenr\=o HIOKI$^{\:}$}\footnote{E-mail address:
\tt hioki@ias.tokushima-u.ac.jp}
\end{center}

\vspace*{0.7cm}
\centerline{\sl Institute of Theoretical Physics,\ 
University of Tokushima}

\vskip 0.14cm
\centerline{\sl Tokushima 770-8502, JAPAN}

\vspace*{3.5cm}
\centerline{ABSTRACT}

\vspace*{0.6cm}
\baselineskip=20pt plus 0.1pt minus 0.1pt
Current status of Top/QCD studies at linear colliders (LC)
is briefly viewed, classifying topics into two categories:
those within the standard model and those beyond the standard
model.

\vspace*{0.4cm} \vfill

\newpage
\renewcommand{\thefootnote}{\sharp\arabic{footnote}}
\pagestyle{plain} \setcounter{footnote}{0}
\baselineskip=21.0pt plus 0.2pt minus 0.1pt
First I would like to classify the topics on Top/QCD into two
categories: those within the SM (standard model) and those beyond
the SM. From this point of view, QCD is main part of the standard
model, while the top-quark could join both part since we do not know
it definitely yet whether it is a standard quark or not. At this
workshop, six talks were presented in Top/QCD session, and in
addition two related talks were given in $\gamma\gamma$ session.
The former six talks were all on the top quark in the SM and/or QCD,
which are therefore summarized in the SM part although their studies
cannot be totally independent of new-physics search, while the latter
two talks were on studying anomalous top-quark interactions, which
are put in the beyond-the-SM part.

\vskip 0.3cm \noindent
{\bf 1. Within the SM}


What do we need for more precise tests of the SM?
Important parameters are $m_t$ and $\alpha_{\rm QCD}$. At present,
they are known with the following uncertainties:
\begin{equation}
{\mit\Delta}m_t^{\rm exp}=4.3\ {\rm GeV},\ \ \ \ \
{\mit\Delta}\alpha_{\rm QCD}^{\rm exp}(M_Z)=0.003. \label{error}
\end{equation}

Why do we have to know them more precisely? Weiglein gave a talk
on this theme \cite{Weiglein}. Knowing them is important for \\
$\bullet$ EW precision tests and Higgs-boson mass prediction \\
$\bullet$ Testing the idea of Grand Unification \\
Let me take the most precise EW formula on the $M_W$-$M_Z$ relation
as an example:
\begin{equation}
M_W^2(1-M_W^2/M_Z^2)=\frac{\pi\alpha}{\sqrt{2}G_F}
(1+{\mit\Delta}r),
\end{equation}
where ${\mit\Delta}r$ expresses all the higher order corrections,
and it is presently known at complete two-loop plus leading
three-loop level. If $m_t$ is measured with ${\mit\Delta}m_t
=1.5$ GeV (0.1 GeV), $M_W$ can be calculated with ${\mit\Delta}M_W
=9$ MeV (1 MeV).
On the other hand, LC is expected to measure $M_W$ with about
7 MeV uncertainty, which realizes an extremely high-precision
test of the SM. Of course, they will also give a strong constraint
on SUSY models and others.


Then how can we measure $m_t$ at LC? One effective way is to use
the threshold behavior of $\sigma(e\bar{e}\to t\bar{t})$, on which
a talk was given by Steinhauser \cite{Steinhauser}. In the
threshold region, $t\bar{t}$ CM frame is a kind of $t\bar{t}$
rest frame. This enables us to take a non-relativistic QCD
approximation. Calculating the QCD potential within this
approximation and comparing thus-calculated cross section
with experimental data, it was shown that we could expect
${\mit\Delta}m_t=$80 MeV. This technique is also applicable to
$b$- and $c$-quark systems.


On the other hand, we need QCD higher order corrections in order
to measure the strong coupling constant $\alpha_{\rm QCD}$.
Weinzierl gave a talk \cite{Weinzierl} on their calculations for
NNLO corrections to
\[
e\bar{e}\ \to\ q\bar{q}\ \to\ 3\,jets.
\]
It is not hard to imagine they are quite tough work: what have
to be computed are not only the two-loop amplitudes of the process,
but also one-loop 4-jet amplitudes plus Born 5-jet amplitudes
to cancel the IR divergence. The theoretical error in the
extraction of $\alpha_{\rm QCD}$ will be thereby reduced down to
1 \%.


Measuring $t\bar{t}H$ coupling is also a significant work, on
which Besson gave a talk \cite{Besson}. This will enable a test
of the SM-vertex, i.e., a coupling proportional to $m_t$. The
Feynman diagrams are those of $H$ emissions from $t$ or $\bar{t}$
in $e\bar{e}\to t\bar{t}$. After a careful study of possible
backgrounds, they are expecting
\begin{eqnarray}
&&{\mit\Delta}g_{t\bar{t}H}/g_{t\bar{t}H}
\lesim\ 10\,\%\ \ \ \ {\rm for}\ m_H \leq 190\,{\rm GeV} \\
&&\phantom{{\mit\Delta}g_{t\bar{t}H}/g_{t\bar{t}H}}
\sim \phantom{1}\mbox{5-6}\,\%\ \ {\rm for}\ m_H \simeq
120\,{\rm GeV}
\end{eqnarray}
for $\sqrt{s}=800$ GeV.


LC can also offer a good opportunity for a traditional hadron
physics ``Pomeron" (and ``Odderon"), which is the theme of
Wallon's talk \cite{Wallon}. Pomeron is something exchanged in
$NN$ forward scattering, that has vacuum quantum number. In
terms of QCD, it is interpreted as the exchange of a bound
state of two gluons. Wallon proposes to use $J/\psi$, $\rho$
productions in the two-photon process in $e\bar{e}$ collisions,
which will work as a test of soft IR part of QCD.


In order to perform those measurements/analyses with small
systematic errors, careful studies of Beam spread, Beam
strahlung, and Initial-state radiation are required. This was
discussed in the talk by Boogert \cite{Boogert}. They are
trying to parameterize those effects into one function, $p(x)$,
with which we can calculate cross sections as
\begin{equation}
\sigma(\sqrt{s})=\int_0^1\!dx\,p(x) \sigma'(x\sqrt{s}),
\end{equation}
where $x$ is a energy fraction. According to their fit results
using $m_t=$175 GeV and $\alpha_{\rm QCD}=$0.118, the following
systematic shifts were observed:
\begin{equation}
{\mit\Delta}m_t=-48\ {\rm MeV},\ \ \ \ \
{\mit\Delta}\alpha_{\rm QCD}(M_Z)=-0.0017.
\end{equation}

\vskip 0.3cm \noindent
{\bf 2. Beyond the SM}

The top-quark mass is even close to the EW breaking scale. This
fact may be an indication that the top-quark possesses some
information which the other quarks do not have. This consideration
leads us to tests of top-quark couplings. For this purpose, one
good signal will be $C\!P$ violation, since $C\!P$ violation in
the SM top-quark couplings occurs at three-loop level and therefore
negligible.

\noindent
{\bf a. $\mib{e}\bar{\mib{e}}$ collision}

Since $t\bar{t}$ is produced via $s$-channel $\gamma/Z$ exchanges
and $m_e$ is negligible comparing to $\sqrt{s}$, initial
$\ket{e\bar{e}}$ must be always $C\!P$ even. Therefore, we have
to construct $C\!P$-odd observables from final-state products :
\[
e\bar{e}\ \to\ t\bar{t}\ \to\ \ell^{\pm}X,\ \ell^+\ell^-X,\ bX.
\]
Many authors have studied and shown that we have good
chances to detect anomalous effects from $t\bar{t}\gamma/Z$
and/or $tbW$ couplings unless the size of them is
extremely small, and for those studies the
use of polarized beams is quite effective.

\noindent
{\bf b. ${\mib\gamma\gamma}$ collision}

In this case, initial states can be $C\!P$ odd, which means we
can make $C\!P$-violating quantities without relying on the
final top-quark (and their decay products) distributions. We
could also study anomalous top-Higgs couplings. Related talks
were given by Asakawa and Hioki in $\gamma\gamma$
session \cite{Asakawa,Hioki}. Although $\sqrt{s_{\gamma\gamma}}$
is not a constant and consequently necessary calculations are
more complicated, similar precision can be expected for some
coupling determination by adjusting initial beam polarizations.

Those $e\bar{e}$ and $\gamma\gamma$ colliders will work
complementary to each other. Let me remind the readers of two
useful tools for performing non-SM interaction analyses through
these processes. One is a decoupling theorem on the final-lepton
angular distributions \cite{Grzadkowski:2002gt} and the other
is the optimal-observable analysis \cite{optimal}. The former
decoupling theorem says that the lepton angular distribution
is free from anomalous top-decay interactions whatever type of
couplings we assume and it holds not only for the final lepton
in $e\bar{e},\ \gamma\gamma$ collisions but also for the one
in single top-quark productions at hadron colliders. On the
other hand, the latter procedure tells us how to determine
several unknown parameters altogether with the least
statistical uncertainty.

Finally let me ask all of us again ``Is the top-quark a
standard quark?". Many people will answer ``Yes, I believe so",
I suppose,  considering the great success of the EW precision
analyses. However, ``No" must be a much much more exciting
answer (at least we could write many papers!). Anyway we hope
LC will be able to give us a clear answer in the near future.
Merci beaucoup!

\vskip 1cm
 \def\NCA{\em Nuovo Cimento}
 \def\NIM{\em Nucl. Instrum. Methods}
 \def\NIMA{{\em Nucl. Instrum. Methods} {\bf A}}
 \def\NPB{{\em Nucl. Phys.} {\bf B}}
 \def\PLB{{\em Phys. Lett.} {\bf B}}
 \def\PRL{\em Phys. Rev. Lett.}
 \def\PRD{{\em Phys. Rev.} {\bf D}}
 \def\EPJ{{\em Eur. Phys. J.} {\bf C}}
 \def\ZFP{{\em Z. Phys.} {\bf C}}

\end{document}